\documentclass[aps,pra,twocolumn,superscriptaddress,longbibliography,preprintnumbers,
nofootinbib, amsmath,amssymb]{revtex4-1}

\usepackage{graphicx}
\usepackage{dcolumn}
\usepackage{bm}
\usepackage{hyperref}
\usepackage{amsmath}
\usepackage{amssymb}
\usepackage{longtable}
\usepackage{color}

\begin{document}


\title{\textbf{Guide to nuclear polarization in muonic atoms.} 
}%

\author{Mikhail Gorchtein}
\email{gorshtey@uni-mainz.de}
\affiliation{
Institut f\"ur Kernphysik, Johannes Gutenberg-Universit\"at Mainz, 55128 Mainz, Germany
}
\affiliation{
PRISMA$^+$ Cluster of Excellence, Johannes Gutenberg-Universit\"at Mainz, 55128 Mainz, Germany
}
\date{\today}

\begin{abstract}
This article deals with the so-called nuclear polarization correction to the 1S-levels in light to intermediate muonic atoms. An easy to use recipe to compute it is given. The calculation includes the effect of the nucleon polarization, i.e. the contribution from inelastic states in the hadronic range, and Coulomb corrections beyond the leading logarithm approximation to both nuclear and nucleon polarization. 
\end{abstract}

\maketitle


Precise nuclear radii play an important role in low-energy tests of the Standard Model (SM) in the quark sector. The pertinent parameters of the SM Lagrangian, such as charges and mixing angles, refer to quarks, whereas the experiments are performed with hadrons, their bound states. The effects of this binding have to be removed to analyze high-precision measurements in terms of the fundamental SM parameters. Because of the nonperturbative nature of the strong force at low energies, such modifications may be large and difficult to compute accurately. It is thus common to identify certain parameters that are well-defined theoretically yet accessible experimentally, which can be taken from the data if it guarantees better precision. 

Nuclear charge radii belong to such parameters that are widely used in many precision tests. The extraction of Cabibbo's 2-flavor quark-mixing angle $\theta_C$, $V_{ud}=\cos\theta_C$, from superallowed nuclear beta decays at the 0.01\% level crucially depends upon precise charge radii of the participating nuclei \cite{Gorchtein:2023naa}. They enter through Coulomb corrections to the statistical rate function $f$ \cite{Seng:2023cgl}, and via the isospin-breaking correction $\delta_C$ which can be benchmarked with combinations of nuclear radii across the superallowed isotriplet \cite{Seng:2022epj}. Nuclear radii enter the calculations of these quantities as external input, and their uncertainties directly affect the uncertainties of $f$, $\delta_C$ and hence $V_{ud}$. The impact of a precise charge radius on the $V_{ud}$ extraction is illustrated by the recent Ref.~\cite{Plattner:2023fmu,Al26-letter} where a 0.5\% determination of the radius of the ${}^{26m}$Al isomer results in a $1-2\sigma$ shift in $\delta_C$ and in $V_{ud}$ extracted from its superallowed decay rate. 
Modern ab-initio methods provide a systematically improvable computational framework for evaluating RMS radii and other observables \cite{Ekstrom:2022yea}. At present, the precision of such calculations is of the order 1\% and is insufficient. Therefore, one is forced to resort to radii deduced from experimental data. 
%
%
For stable isotopes, the most precise charge radii are obtained from the x-ray spectra of muonic atoms. The radii of unstable isotopes are obtained via optical isotope shift measurements using as input the reference radii from muonic measurements~\cite{Fricke:2004,Ohayon:2024dwt}. 

To relate the transition frequency between atomic levels, e.g. $1S$-$2P$, to the nuclear radius, a number of corrections need to be computed. QED corrections can be reliably computed by solving the bound-state Schr\"odinger or Dirac equation numerically \cite{Borie:1982ax}. Perturbation theory is well established for atomic calculations (at least for not too heavy atoms). Nonperturbative strong interaction effects also enter here, but at a higher order in $Z\alpha$: 
%
If two energetic photons are exchanged between the muon and the nucleus, the latter can be excited (polarized) and then deexcited by successive interactions. This effect is generally called (leading-order) nuclear polarization (NP). 

In the past few decades, great amount of theoretical work has been dedicated to NP in the lightest hydrogen-like muonic systems (see the recent review~\cite{Pachucki:2022tgl} that contains a comprehensive list of references). This activity was motivated by a technological leap in the laser spectroscopy of muonic H, D,${}^{3,4}$He$^+$ \cite{Antognini:2013txn,CREMA:2016idx,Krauth:2021foz,CREMA:2023blf}.
For heavier muonic atoms, the estimates of NP stem from the 1970's work by Rinker and Speth \cite{Rinker:1978kh}. Although there has been an active discussion in the literature in those years~\cite{Ericson:1972nhh,Bernabeu:1973uf,Rinker:1976en,Friar:1977cf,Bernabeu:1982qy,Rosenfelder:1983aq}, it was mostly dedicated to specific case studies, such as muonic helium. For the $S$ states of other light and medium-light muonic atoms, the NP estimates from Ref.~\cite{Rinker:1978kh} still count as the state-of-the-art and are used to extract nuclear radii from muonic measurements to this date. NP estimates may be explicitly quoted \cite{Fricke:1995zz,Fricke:2004}, or not quoted and used implicitly \cite{Angeli:2013epw}. On the other hand, NP estimates may vary significantly in the literature, even if the calculations are essentially done within the same approach.  For instance, entries for NP for Fe, Co, Ni, Cu, Zn isotopes in Refs.~\cite{Fricke:2004,Shera:1976zz} differ by 30-40\%. 
The interest in the NP correction to the energy levels in heavy atoms persists also in the recent years, see, e.g., Refs.~\cite{Flambaum:2021yyz,Yerokhin:2023dxj,Valuev:2024rxq}.

This letter is dedicated to collecting available information and proposing a simple recipe to compute leading-order nuclear polarization in medium-light nuclei. Most formulas have appeared in the literature already, hence the title: a hitchhiker's guide. The complete road map leading to the final results is laid out so that any practitioner can obtain the NP estimate her/himself. 
A review of the theory of muonic atoms (not limited to light systems) by Borie and Rinker~\cite{Borie:1982ax} includes a detailed discussion of NP but dates back to 1982 and lacks recent developments.
Future plans for improved measurements with nuclei heavier than helium are put forward by several collaborations~\cite{Ohayon:2023hze,Wauters:2021cze,MuSEUM:2023jww,Saito:2022fwi,Adamczak:2022jbo} and an update of NP which would comprise all currently available knowledge and a robust uncertainty estimate is highly desirable. This work is intended as a first step in this direction. Anticipating, I find the effects of the inelastic contributions in the hadronic range, which I coin "nucleon polarization" (nP), non-negligible. It is by now well studied and included in NP for the lightest systems~\cite{Pachucki:2022tgl}, but is  missing in all atoms heavier than helium. Starting from calcium, this correction is of the same size as the experimental precision, so its inclusion is mandatory. 


%

At the leading order in $Z\alpha$, 
NP is given by the following one-loop integral~\cite{Bernabeu:1973uf,Carlson:2011zd},
\begin{align}
    \Delta E_{n\ell}&=\frac{8\alpha^2m}{i\pi}|\phi_{n\ell}(0)|^2\label{eq:TPEloop}\\
    &\times\int d^4q\frac{(q^2-\nu^2)T_2-(q^2+2\nu^2)T_1}{q^4(q^4-4m^2\nu^2)}\nonumber
\end{align}
with $|\phi_{n\ell}(0)|^2=(Z\alpha m_r/n)^3/\pi\delta_{\ell 0}$ the squared atomic WF at the origin, and $m_r=mM/(m+M)$ the reduced mass for the lepton and nucleus masses $m,M$, respectively. The spin-independent forward Compton amplitudes $T_{1,2}(\nu,q^2)$, functions of the energy $\nu=(p q)/M$ and the virtuality $q^2<0$ of the loop photons, encompass all the information about the nuclear structure. 

The dispersion approach is well suited for evaluating the loop integral \eqref{eq:TPEloop}. The imaginary part of the Compton amplitudes is related to the inelastic structure functions $F_{1,2}$,
\begin{align}
    \text{Im}\,T_1(\nu,q^2)&=\frac{1}{4M}F_1(\nu,q^2)\nonumber\\
\text{Im}\,T_2(\nu,q^2)&=\frac{1}{4\nu}F_2(\nu,q^2).
\end{align}
The latter can be obtained from abundant experimental data on real and virtual photoabsorption. The real part of $T_{1,2}$ is obtained from a dispersion representation
which permits an evaluation of the loop integral analytically, and one is left with a twofold integral over the photoabsorption data (i.e. $F_{1,2}$) over the energy and $q^2$ with the known kinematical weighting (see Eq.~(16) in Ref.~\cite{Carlson:2011zd}). The integral over the hadronic range $\nu\geq\nu_\pi$ ($\nu_\pi\approx140$ MeV denotes the pion production threshold) must be treated in a fully relativistic manner. This result can easily be extended to heavier nuclei. The total photoabsorption on the proton and neutron is essentially the same, suggesting that for a generic nucleus the hadronic contribution should roughly scale with the atomic mass number $A$. Studies of this scaling and deviations therefrom $A\to A_\text{eff}<A$, generally coined shadowing in the literature, have been performed on a variety of nuclei. In the resonance region, it was found that $A_\text{eff}\approx A$, such that the integrated cross section is largely unaffected~\cite{Muccifora:1998ct}. At high energies, however, shadowing leads to a more significant suppression: e.g. for lead $A_\text{eff}\approx 0.6A$~\cite{Bianchi:1999gs}. Since the integrand is strongly weighted at lower energies, I take $A_\text{eff}\approx A$. 
The nP correction to the $2S$ level in $\mu D$ amounts to~\cite{Carlson:2013xea}
\begin{equation}
    \left[\Delta E_{2S}^\mathrm{hadr}
    \right]_{\mu D}=-28(2)\,\mu{\rm eV}.\label{eq:muD}
\end{equation}

The respective contribution to the $nS$ level in a muonic atom $\mu A$ will then read
as
\begin{align}
   \left[\Delta E_{nS}^\mathrm{nP}
    \right]_{\mu A}&=-28(2)\,\mu{\rm eV} \frac{|\phi_{nS}^{\mu A}(0)|^2}{|\phi_{2S}^{\mu D}(0)|^2}\frac{A}{2}.\label{eq:nP}
\end{align}
Note that this contribution has traditionally been neglected in all atoms heavier than helium. {The authors of Ref.~\cite{Pachucki:2018yxe} proposed including nP in a similar manner but opted to include the elastic contribution. In the approach used here, this would lead to a double-counting, so I only account for the inelastic part.}

Next, I turn to the nuclear range $\nu<\nu_\pi$. 
Although the same entirely data-driven approach may also be used here, the resulting estimates are plagued by large uncertainties~\cite{Carlson:2013xea,Carlson:2016cii} that can only partially be cured by the approximate finite-energy sum rules~\cite{Gorchtein:2015eoa}. Instead, a hybrid approach which combines the data input in the hadronic range with the microscopic nuclear theory at low energies is more reliable~\cite{Acharya:2020bxf}. Such calculations need to be performed case by case. Here, I aim for a more synthetic calculation applicable for a range of nuclei. 

To proceed, I note that the bulk of nuclear excitations resides at $\nu_N\leq35$ MeV, rather well separated from the hadronic range. Because nuclear excitation energies are small compared to the nucleon mass, the nuclear part of the integral can be treated nonrelativistically. Here, I follow Refs.~\cite{Friar:1977cf,Rosenfelder:1983aq}. The most important part of the nuclear response is due to the longitudinal response function,
\begin{equation}
    \Delta E_{nS}^{NP}=-8\alpha^2|\phi_{nS}(0)|^2\int_0^\infty\frac{d\mathbf{q}}{\mathbf{q}^2}\int_0^\infty\frac{d\nu S_L(\nu,\mathbf{q})}{\nu+\mathbf{q}^2/2m},\label{eq:NPlongRF}
\end{equation}
where the longitudinal response function $S_L$ is taken in the retarded dipole approximation,
\begin{equation}
S_L(\nu,\mathbf{q})=\mathbf{q}^2\frac{\sigma_\gamma(\nu)}{4\pi^2\alpha\nu}F^2(\mathbf{q}),
\end{equation}
with $\sigma_{\gamma}(\nu)$ the total photoabsorption cross section in the nuclear range. 
The electric dipole polarizability is given by its $-2$ moment,
\begin{equation}
    \alpha_{E1}=\frac{1}{2\pi^2}\int\frac{d\nu}{\nu^2}\sigma_\gamma(\nu).
\end{equation}
The nuclear form factor is taken in Gaussian form $F(\mathbf{q})=exp(-\mathbf{q}^2R_{ch}^2/6)$. 
The $\mathbf{q}$-integral can be taken analytically. The $\nu$-integral has $\nu^{-3/2}$ weighting~\cite{Friar:1977cf} but can be approximated by that with $\nu^{-2}$ weighting since nuclear photoabsorption is strongly peaked at an energy $\sim15-25$ MeV, depending on the nucleus. 
I arrive at
\begin{align}
    \Delta E_{nS}^\text{NP}=-2\pi\alpha|\phi_{nS}(0)|^2 \alpha_{E1}\sqrt{2m\bar\nu}\,e^{\beta^2(\bar\nu)}\mathrm{Erfc}(\beta(\bar\nu)),\label{eq:NPresult}
\end{align}
with $\beta(x)=\sqrt{2mx/3}R_\text{ch}$, $R_{ch}$\footnote{Note added in proof: the author is grateful to T. Egert and S. Bacca for pointing to a typo in the original version where $\beta(x)=2mx/3 R_\text{ch}^2$ was used.} standing for the respective nuclear charge radius, and $\mathrm{Erfc}$ is the complementary error function. Ref.~\cite{Ji:2018ozm} represents NP as a Taylor expansion in powers of $\eta\sim\beta(\bar\nu)$ and accounts for first few terms, an approximation appropriate for lightest atoms. The result of Eq.~\ref{eq:NPresult} avoids such an expansion keeping the full $\beta$ dependence, which permits to extend the description to larger atomic numbers.

As a check of the model dependence, the above integral was evaluated with the form factor corresponding to the homogeneously charged sphere distribution, $F(\mathbf{q})=3J_1(qR)/qR$ with $R=\sqrt{5/3}R_{ch}$ the radius of the sphere. 
The difference between the two never exceeds 1\%. 
%

The dipole polarizability is an external input for which
I use the empirical scaling formula obtained from a fit from oxygen to lead~\cite{Steiner:2004fi,vonNeumann-Cosel:2015jta,Orce:2015lwa,Orce:2023qtm}
\begin{equation}
    \alpha_{E1}=\frac{0.0518\,\mathrm{MeV~fm}^3 A^2}{S_v(A^{1/3}-\kappa)},\label{eq:pol_fit}
\end{equation}
with $S_v=27.3(8)$~MeV and $\kappa=1.69(6)$. 
For lighter elements I use the values from Ref.~\cite{Ahrens:1975rq}. Since that Ref. does not cover ${}^{10}$B or ${}^{14}$N I extrapolate  from the measured polarizability of ${}^{12}$C assuming for simplicity the $A^{5/3}$ scaling law. 
Within the range of validity of the fit of Eq.\eqref{eq:pol_fit} (oxygen and above) the uncertainty always stays well below 10\%. To take into account that individual polarizabilities may deviate from the fit by more than $1\sigma$, I assign a generic 10\% uncertainty on the normalization of $\alpha_{E1}$ in the entire range, and use the central values $S_v=27.3$~MeV and $\kappa=1.69$. 

The value of the mean excitation energy $\bar\nu$ is also deduced from the moments of the photoabsorption cross section $\sigma_{-n}=\int d\nu\sigma_{\gamma}(\nu)/\nu^n$ for $n=0,1,2$. I define 
\begin{equation}
\bar\nu={\sigma_{-1}}/{\sigma_{-2}}
.\label{eq:av_energy}
\end{equation}
The values of $\sigma_{-n}$ are taken from \cite{Ahrens:1975rq,Berman:1975tt}. In case the entry is missing, the value for the closest neighbor element from \cite{Berman:1975tt} is adopted. Since $\bar\nu_N$ changes very little between nearby elements, the associated uncertainty does not exceed 1-2\%, well below other sources of uncertainty. 

Eqs.\eqref{eq:nP},\eqref{eq:NPresult},\eqref{eq:pol_fit},\eqref{eq:av_energy} represent the result at the leading order in $Z\alpha$.

It is well known that even for low $Z$ the next-to-leading order corrections are non-negligible. 
The approximation scheme underlying Eq.\eqref{eq:TPEloop} assumes that (i) the atomic size is much larger than the nuclear one, $(Z\alpha m_r)^{-1}\gg R_{ch}$; (ii) nuclear excitations lie at energies $\nu_N$ that are much larger than atomic ones, $\nu_N\gg (Z\alpha)^2m_r/2$. To extend the validity of the calculation, one should include the higher-order corrections in the two specified expansion parameters,  $\epsilon_1=Z\alpha m_r R_{ch}$ and $\epsilon_2=(Z\alpha)^2m_r/2\nu_N$. 

The reduction factor $F_R$ accounts for the variation of the atomic 1S-wave function squared $\sim exp(-2Z\alpha m_r r)$ over the nucleus volume. The nuclear charge distribution is taken for simplicity in the Gaussian form $\sim exp(-3r^2/2R_{ch}^2)$. This gives 
\begin{equation}
    F_R=\int_0^\infty r^2dr e^{-2Z\alpha m_r r}\,\frac{3\sqrt{6}}{\sqrt{\pi}R_{ch}^3}e^{-\frac{3r^2}{2R_{ch}^2}},\label{eq:RedFac}
\end{equation}
and it quantifies the corrections in the expansion parameter $\epsilon_1$. This correction accounts for the spatial distribution of the probability for the nucleus to be polarized by the orbiting muon. Since the strong interaction responsible for nuclear transitions is short-range, the muon should be on top of the active nucleons. This correction applies to both NP and nP. 
To estimate the uncertainty, I also compute $R$ using the homogeneous sphere distribution corresponding to the same charge radius,
\begin{equation}
 F'_R=\int_0^{R_{sph}} \frac{3r^2dr}{{R_{sph}}^3} e^{-2Z\alpha m_r r},\quad {R_{sph}}=\sqrt{\frac{5}{3}}R_{ch}\label{eq:RedFac2}
\end{equation}

\begin{figure}[h]
\includegraphics[width=1.0\linewidth]{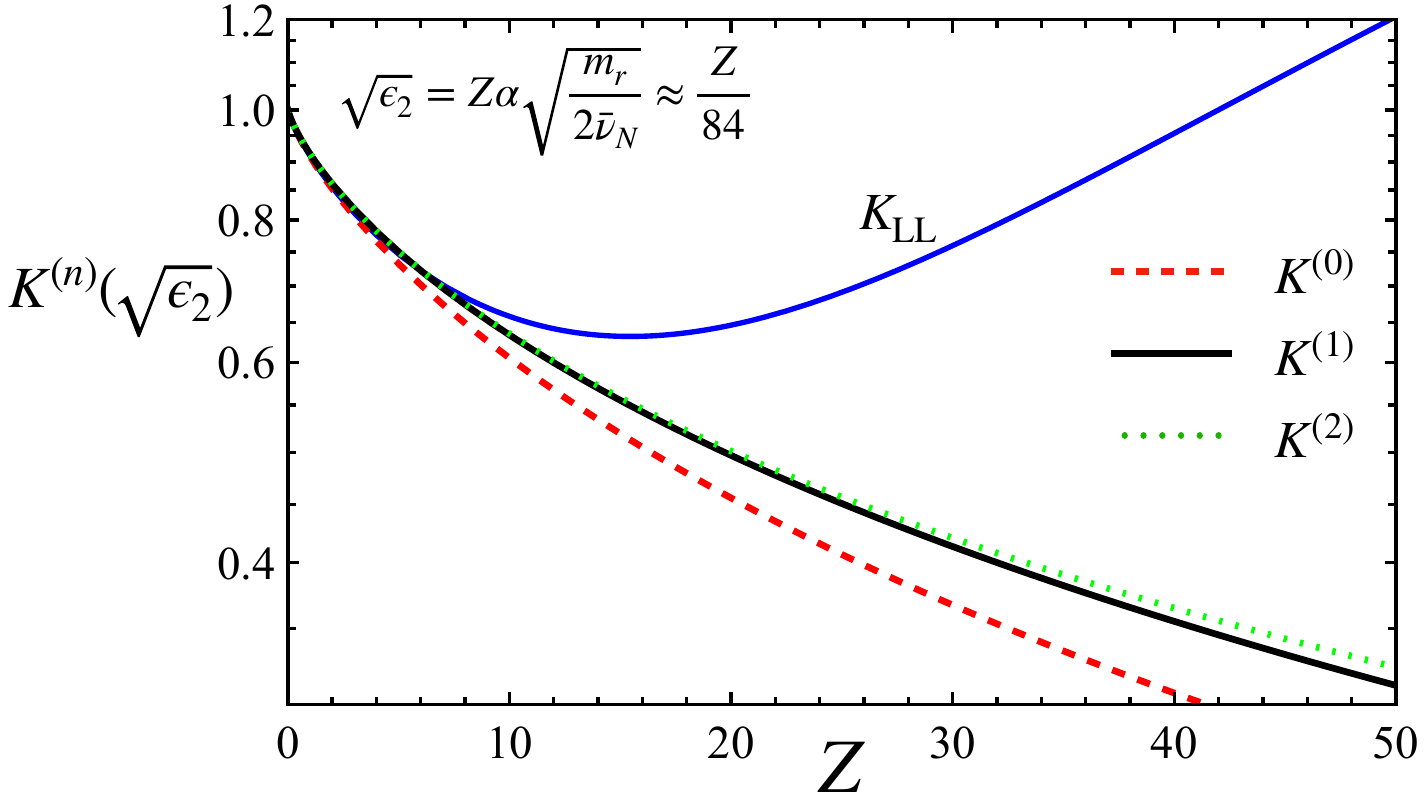}
\caption{\label{fig:CDcomparison} Radial Coulomb integrals $K^{(0)}$ (red dashed curve), $K^{(1)}$ (solid black curve), and $K^{(2)}$ (dotted green curve) in comparison with the leading-logarithm approximation (solid blue curve).}
\end{figure}

To include higher orders in $\epsilon_2$, I account for the Coulomb distortion of the muon propagator inside the loop, following Ref.~\cite{Friar:1977cf} (see also details reported in Ref.~\cite{Ji:2018ozm}). Coulomb interaction is described by the point Coulomb radial Green's function defined by 
\begin{equation}
    \left[\frac{1}{2m_r}\frac{d^2}{dr^2}-\frac{l(l+1)}{2m_r r^2}+\frac{Z\alpha}{r}+E\right]g_l(E,r,r')=\delta(r-r').\label{eq:GF}
\end{equation}
In the unretarded dipole approximation, the muon Green's function should be taken for $l=1$ and for $E=-\nu_N$~\cite{Friar:1977cf}. The task is reduced to the following radial integral:
\begin{equation}
    K=-\sqrt\frac{\nu_N}{2m_r}\int\limits_0^\infty dr\int\limits_0^\infty dr'\phi_{nS}(r)\frac{g_1(-\nu_N,r,r')}{rr'}\phi_{nS}(r').
\end{equation}
Such integrals have been evaluated in the general case in the literature~\cite{Swainson_1991}. The integral at hand is a special case of the integral $K_{\mu_1\mu_2}^{\nu\lambda}(p_1,p_2,\omega)$ defined in Eq.(5.1) of Ref.~\cite{Swainson_1991}\footnote{Note that Ref.~\cite{Swainson_1991} uses the definition $g=-2m_r g/rr'$ with respect to that used here.}. For the $1S$ states, the values of the parameters should be chosen as $\mu_{1,2}=0$, $\lambda=1$, $p_{1,2}=\frac{Z\alpha}{2}\sqrt{\frac{2m_r}{\nu_N}}$ and $\omega=\sqrt{2m_r\nu_N}$.
Using the representation in terms of the Gauss hypergeometric function in Eq.~(5.7) of Ref.~\cite{Swainson_1991}, I find 
\begin{align}
    K=\frac{2}{9}\dfrac{
    1
    }{(1+p)^4}\sum_{k=0}\frac{\Gamma(k+4)}{k!(2+k-p)}\,\left[{}_2F_1(2,-k;4;\frac{2}{1+p})\right]^2.
\end{align}
The sum can be performed analytically in a closed form if the $p$-dependence under the sum is only kept in the hypergeometric function. 
To proceed, I Taylor-expand the denominator,
\begin{equation}
    \frac{1}{2+k-p}=\frac{1}{2+k}\sum_{m=0}\left(\frac{p}{2+k}\right)^m,\label{eq:series}
\end{equation}
which is justified because $p=\sqrt{\epsilon_2}<1$ in the approximation scheme used here. Each term in the expansion can be evaluated analytically, e.g. using Mathematica. Denoting with $K^{(n)}$ the result of wrapping the series in Eq.\eqref{eq:series} at the power $p^n$ (i.e., $K^{(n)}=\dots\sum_{m=0}^n\frac{p^m}{(2+k)^{m+1}} \dots$) and introducing a shorthand for the often recurring combination, $\xi=\frac{p-1}{p+1}$, I find
\begin{align}
    K^{(0)}&=1+2p\ln(1+\xi)\label{eq:K0}
    -p^2\left[\text{Li}_2\left(-\xi\right)+\frac{\pi^2}{12}\right],\\
    K^{(1)}&=1+2p\ln(1-\xi^2)-p+p^2\left(1-{2}\text{Li}_2\left(\xi^2\right)\right)\label{eq:K1}\\
    &+p^3\left[\text{Li}_3\left(\xi\right)-\frac{3}{2}\text{Li}_3\left(\xi^2\right)+\ln(1+\xi)-\frac{\zeta(3)}{2}\right],\nonumber\\
K^{(2)}&=1+2p\ln(1-\xi^2)-p\label{eq:K2}+p^2\left[1-{2}\text{Li}_2(\xi^2)\right]\\
    &+p^3\left[\text{Li}_3(\xi)-\frac{3}{2}\text{Li}_3(\xi^2)+\ln(1+\xi)+\frac{\zeta(3)}{2}\right]\nonumber\\
    &+p^4\left[\text{Li}_4(\xi)-\frac{1}{2}\text{Li}_4(\xi^2)-    \text{Li}_2\left(-\xi\right)-
    \frac{\pi^2}{12}-\frac{\pi^4}{180}\right]\nonumber
\end{align}
with $Li_n$ denoting the polylogarithm and $\zeta$ the Riemann zeta function. Note that the result of Ref.~\cite{Friar:1977cf} widely adopted in light muonic atoms corresponds to only keeping the leading logarithm,
\begin{equation}
    K_\mathrm{LL}=1+2p\ln 2p
\end{equation} 
in Eq.\eqref{eq:K0}. In Fig.\ref{fig:CDcomparison} I show the effect of including higher orders up to $p^4$. All curves are seen to agree nicely below $Z=5$ but the leading-logarithm result starts to deviate from $K^{(1)}$ above that value. For the numerical estimates, I will use $K^{(1)}(\sqrt{\epsilon_2})$ for the central value and half the difference, $(K^{(1)}-K^{(0)})/2$, as an uncertainty estimate. This is conservative because the higher-order result $K^{(2)}$ only differs from $K^{(1)}$ very little, as seen in Fig.\ref{fig:CDcomparison}. It would be interesting to compare these results to the recently considered three-photon exchange correction to NP~\cite{Pachucki:2018yxe}, especially in view of the fact that the leading-order approximation was found here to be ill-behaved already for moderate atomic numbers.

This brings me to the final expression that can be used for numerical estimates:
\begin{align}
    \Delta E^\text{TOT}_{nS}&=\Delta E^\text{NP}_{nS}\,F_R(\epsilon_1)\,K^{(1)}(\sqrt{\epsilon_2})\nonumber\\
    &+\Delta E^\text{nP}_{nS}\,F_R(\epsilon_1)K^{(1)}(\sqrt{\epsilon_2^n}),
\end{align}
with $\Delta E^{NP}_{nS}$ as given in Eq.\eqref{eq:NPresult}, $\Delta E^{nP}_{nS}$ as given in Eq.\eqref{eq:nP}, $F_R$ from Eq.\eqref{eq:RedFac} and $K^{(1)}$ from Eq.\eqref{eq:K1}.
The Coulomb correction to nP is evaluated at $\epsilon_2^n=(Z\alpha)^2m_r/2\nu_n$ with $\nu_n\approx500$ MeV the mean excitation energy in the hadronic range.

The overall uncertainty is composed as follows. For NP: (i) 10\% uncertainty on $\alpha_{E1}$; (ii) uncertainty of the reduction factor $F_R$, conservatively estimated as 100\% difference between Eqs.\eqref{eq:RedFac} and \eqref{eq:RedFac2}; (iii) uncertainty on the Coulomb correction obtained as half the difference of Eqs.\eqref{eq:K1} and \eqref{eq:K0}.
For nP, the uncertainty results by combining the latter two uncertainties with the 10\% on the input in Eq.\eqref{eq:muD}. I add the individual uncertainties in quadrature. This treatment rests on an assumption that all uncertainties are Gaussian and independent. I note here that the finite nuclear size effects are accounted for twice: in Eq.\eqref{eq:NPresult} via the $\beta$-dependence, and in the atom-nucleus volume overlap $F_R$, while neglected in the Coulomb correction. This apparently leads to a double counting~\cite{Rosenfelder:1983aq}. Removing this double counting and including the finite nuclear size in the Coulomb correction would lead to {\it larger} values of Npol. To stay conservative, I assign the total Npol an asymmetric uncertainty, doubling it towards larger values.

\begin{table*}[ht]
\caption{\label{tab:table1}%
Nuclear dipole polarizability in units of fm$^3$, 
nuclear and nucleon polarization contributions to the $2p_{3/2}$-$1s_{1/2}$ transition in selected light and intermediate muonic atoms in units of eV, and the sum of the two in comparison with the respective entries in Ref.~\cite{Fricke:2004}. The three uncertainties refer to the polarizability, $F_R$ and $K^{(1)}$, respectively. NP to $2p$ states is ignored as it is much smaller than the uncertainty. The superscript $^a$ ($^{a*}$ ) at the value of $\alpha_{E1}$ for light nuclei indicates that it is taken from Ref.~\cite{Ahrens:1975rq} (extrapolated from neighboring nuclei according to the $A^{5/3}$ scaling), while no superscript implies that it is evaluated using the fit of Eq.~\eqref{eq:pol_fit}. 
The last column shows the experimental accuracy of recent and future measurements: the planned QUARTET collaboration measurements~\cite{Ohayon:2023hze,BenPC} ($^{b}$), the recent chlorine isotopes measurement~\cite{Beyer:2025imi} ($^{c}$) and the future RefRad potassium measurements~\cite{Adamczak:2022jbo,MichaelPC}.
}
\begin{ruledtabular}
\renewcommand{\arraystretch}{1.18}
\begin{tabular}{llcrrrrr}
$Z-$Element & $A$ & $\alpha_{E1}$\,(fm$^3$) & $-\Delta E_{1S}^{NP}$ &$-\Delta E_{1S}^{nP}$& Total Npol & Entry in \cite{Fricke:2004} & $\sigma_\text{exp}$
 \\
\hline
$3-$Li& 6 & 0.152(15)$^{a*}$ & 0.146(15)(0)(1) & 0.018(2)(0)(0) & 0.16$_{-2}^{+4}$ & - & 0.1$^b$
\\ 
& 7 & 0.196(20)$^a$ & 0.153(15)(0)(1) &  0.021(2)(0)(0)& 0.17$_{-2}^{+4}$ & - & 0.1$^b$
\\ 
$4-$Be& 9 & 0.192(19)$^a$ & 0.35(3)(0)(0)& 0.063(6)(0)(0) & 0.41$_{-4}^{+8}$ & 1.0(3) & 0.2$^b$
\\ 
$5-$B& 10 & 0.230(23)$^{a*}$ & 0.80(8)(0)(1)& 0.13(1)(0)(0)&  0.93$_{-10}^{+20}$ & 1.0(3) & 0.3$^b$
\\ 
$6-$C& 12 & 0.313(31)$^a$ & 1.7(2)(0)(0) & 0.27(3)(0)(0) & 1.9$_{-2}^{+4}$ & 2.5(7) & 0.4$^b$
\\ 
$7-$N& 14& 0.405(40)$^{a*}$ & 3.0(3)(0)(1) & 0.48(5)(0)(0) & 3.5$_{-3}^{+6}$ & 3.0(9) & 0.5$^b$
\\
$8-$O& 16& 0.580(58)$^a$ & 6.2(6)(1)(1) & 0.79(8)(1)(0) & 7.0$_{-0.7}^{+1.4}$ & 5.0(1.5) & 0.7$^b$
\\
$9-$F& 19& 0.700(70) & 9.7(1.0)(0.1)(0.2) & 1.28(13)(1)(1) & 11.0$_{-1.1}^{+2.2}$ & 9.0(2.7) & 1.0$^b$
\\
${10}-$Ne& 20& 0.741(74)& 12.6(1.3)(0.2)(0.3) & 1.78(18)(2)(1) & 14.3$_{-1.3}^{+2.6}$ & 19(6) & 2.0$^b$
\\
& 22& 0.823(82)& 14.5(1.4)(0.2)(0.3) & 1.98(20)(2)(1) & 17$_{-2}^{+4}$ & 18(5) & 2.0$^b$
\\
${17}-$Cl& 35& 1.47(15)& 81.4(8.1)(2.4)(2.8) & 11.9(1.2)(0.3)(0.1) & 93$_{-9}^{+18}$ & - & 16$^c$
\\
& 37& 1.58(16)& 88(9)(3)(3) & 12.6(1.3)(0.4)(0.1) & 100$_{-10}^{+20}$ & - & 12$^c$
\\
${19}-$K& 39& 1.70(17)& 118(12)(5)(5) & 18(1.8)(0.6)(0.2) & 136$_{-14}^{+28}$ & 119(36) & 10$^d$
\\
& 41& 1.81(18)& 126(13)(5)(6) & 18(1.8)(0.6)(0.2) & 144$_{-15}^{+30}$ & 132(40) & 10$^d$
\\
\end{tabular}
\end{ruledtabular}
\end{table*}

The results of the calculation for selected light elements are shown in Tab.\ref{tab:table1}. Tabs.~
\ref{tab:tablex},\ref{tab:tablexx} 
in the Appendix show these results for all stable isotopes of elements with $3\leq Z\leq41$ along with the respective entries in Ref.~\cite{Fricke:2004}. Generally, a good agreement for the nuclear part is observed, within the errors. This is reassuring, since the input used here differs significantly from that used in  Ref.~\cite{Fricke:2004} for obtaining the radii. Ref.~\cite{Rinker:1978kh} which serves as a basis for those calculations, uses the energy-weighted sum rule ($\sigma_0$) to normalize the NP, rather than the needed $\sigma_{-3/2}$. I use $\sigma_{-2}$ related to the polarizability which is much closer. It has been argued that the polarizability is strongly affected by the low-lying ``pygmy dipole resonance" to which the energy-weighted sum rule has less sensitivity~\cite{Piekarewicz:2010fa,Savran:2013bha}. 
Rather than using the phenomenological approach of \cite{Rinker:1978kh} based on approximating the effective muon-induced potential by a power $r^k$~\cite{Ford:1969rzz}, 
I explicitly account for higher-order corrections in $Z\alpha$ by computing an overlap of the atomic wave functions with the nuclear charge distribution and Coulomb corrections. For the latter, I show that the approximate formulas used for light muonic atoms are ill-suited even for moderate $Z$, and the exact result should be expanded to higher orders.

The bulk of nuclear polarization ($\Delta E^{NP}$) is directly proportional to the nuclear polarizability. 
The fit of Eq.~\eqref{eq:pol_fit} for the latter introduces an uncertainty in the prediction. Evaluating Eq.~\eqref{eq:pol_fit} for several nuclei for which modern nuclear calculations of $\alpha_{E1}$ exist, a comparison reveals  agreement within errors, e.g., 1.75(18) fm$^3$ vs. 1.92(17) fm$^3$ \cite{Fearick:2023lyz} for $^{40}$Ca, and 2.25(23) fm$^3$ vs. 2.07(22) fm$^3$ \cite{Birkhan:2016qkr} for $^{48}$Ca, or slight deviation, as for $^{58}$Ni, 2.93(29) fm$^3$ vs. 3.48(31) fm$^3$ \cite{Brandherm:2024rci}.
In this work, $\alpha_{E1}$ is an external input and it is its connection to $\Delta E^{NP}_{1S}$ that is obtained; changing the value of the former would rescale the latter accordingly. To facilitate this rescaling to the practitioners, I include the value and assumed uncertainty of $\alpha_{E1}$ in Tables \ref{tab:tablex},\ref{tab:tablexx} next to $\Delta E^{NP}_{1S}$. 
For light muonic atoms, agreement within the errors is observed with other older works, e.g. Ref.~\cite{Drake:1985tp} for $\mu\,^9$Be and $\mu\,^{10}$B and Ref.~\cite{Rosenfelder:1983aq} for $\mu\,^{12}$C. The differences can be traced back to a different choice of the nuclear dipole polarizability, neglect of Coulomb corrections or phenomenological ansatzes in those works.
In the case of ${}^{90}$Zr, a comparison can be made with more modern calculations. Ref.~\cite{Valuev:2022tau} gives a range of 1408-1560~eV, to be compared to my estimate of the pure nuclear polarization of $1059_{-182}^{+364}$~eV. 
Note that the nP contribution to 1S state in $\mu{}^{90}$Zr is 166(26)~eV, non-negligible for this comparison. To assess, whether the agreement deteriorates if going to heavier nuclei, I also checked the extreme case of $\mu{}^{208}$Pb: my estimate $3.9_{-1.5}^{+3.0}$~keV agrees with 5.7(6)~keV~\cite{Valuev:2022tau,Sun:2025qll} within the uncertainties. Since for $\mu{}^{208}$Pb the expansion parameters are not small, $\epsilon_1=1.76$ and $\epsilon_2=1.33$, this agreement should not be taken for granted. These large values of the expansion parameters also lead to a very large uncertainty within the very conservative estimate scheme presented here. 
The nP contribution, not included in any of the previous calculations in the shown $Z$ range, is sizable, especially confronted to the accuracy of modern experiments: Indeed, starting from nitrogen, the two are comparable, as seen in Tab.~\ref{tab:table1}. 
Future work will be dedicated to further reducing the readily identified uncertainties, providing predictions for atomic levels beyond the 1S considered here, 
and to removing the double counting of the aforementioned finite size effects. The hitherto neglected effects, e.g. magnetic and higher-multipole excitations, subleading terms, relativistic corrections, will be addressed, as well.

\begin{acknowledgments}
The author is grateful to S. Bacca, T. Egert, J. Erler, B. Ohayon, N. Oreshkina, V. Pascalutsa, and H. Spiesberger for many motivating and enlightening discussions. He furthermore acknowledges support by the EU Horizon 2020 research and innovation program, STRONG-2020 project under grant agreement No 824093, and by the Deutsche Forschungsgemeinschaft (DFG) under grant agreement GO 2604/3-1.
\end{acknowledgments}

\bibliography{apssampapp}

\appendix*
\section{Numerical results}\label{sec:app}
In this appendix I report the numerical values for the nuclear dipole polarizability in units of fm$^3$, nuclear polarization (NP) and nucleon polarization (nP) contributions and the sum of the two (Npol) to the 1s energy level in the muonic atoms units of eV, for the stable isotopes of elements with the atomic number $3\leq Z\leq41$, as obtained from Eqs.\eqref{eq:NPresult}, \eqref{eq:nP}, \eqref{eq:RedFac}, and \eqref{eq:K1}. The three uncertainties refer to the polarizability, $F_R$ and $K^{(3)}$, respectively. NP to $2p$ states is ignored as it is much smaller than the uncertainty. 
The superscript $^a$ at the value of $\alpha_{E1}$ for light nuclei indicates that it is taken from Ref.~\cite{Ahrens:1975rq}, while no superscript implies that it is evaluated using the fit of Eq.~\eqref{eq:pol_fit}. The 
$^{a*}$ superscript indicates an extrapolation from neighboring nuclei according to the $A^{5/3}$ scaling. 
The last two columns show the Npol and the experimental precision quoted in Ref.~\cite{Fricke:2004}.

\begin{table*}[ht]
\caption{\label{tab:tablex}%
Nuclear polarization contribution to 1s states for muonic atoms with $3\leq Z\leq25$, see the text in the appendix for details. 
 }
\begin{ruledtabular}
\renewcommand{\arraystretch}{1.18}
\begin{tabular}{llcrrrrr}
$Z-$Element & $A$ & $\alpha_{E1}$\,(fm$^3$) & $-\Delta E_{1S}^{NP}$ &$-\Delta E_{1S}^{nP}$& Total NP & Entry in \cite{Fricke:2004} & $\sigma_\text{exp}$
 \\
\hline
$3-$Li& 6 & 0.152(15)$^{a*}$ & 0.146(15)(0)(1) & 0.018(2)(0)(0) & 0.16$_{-2}^{+4}$ & - & -
\\ 
& 7 & 0.196(20)$^a$ & 0.153(15)(0)(1) &  0.021(2)(0)(0)& 0.17$_{-2}^{+4}$ & - & -
\\ 
$4-$Be& 9 & 0.192(19)$^a$ & 0.35(3)(0)(0)& 0.063(6)(0)(0) & 0.41$_{-4}^{+8}$ & 1.0(3) & 10
\\ 
$5-$B& 10 & 0.230(23)$^{a*}$ & 0.80(8)(0)(1)& 0.13(1)(0)(0)&  0.93$_{-10}^{+20}$ & 1.0(3) & 7
\\ 
$6-$C& 12 & 0.313(31)$^a$ & 1.7(2)(0)(0) & 0.27(3)(0)(0) & 1.9$_{-2}^{+4}$ & 2.5(7) & 0.5
\\ 
$7-$N& 14& 0.405(40)$^{a*}$ & 3.0(3)(0)(1) & 0.48(5)(0)(0) & 3.5$_{-3}^{+6}$ & 3.0(9) & 5 
\\
$8-$O& 16& 0.580(58)$^a$ & 6.3(0.6)(0.1)(0.1) & 0.79(8)(1)(1) & 7.1$_{-0.7}^{+1.4}$ & 5.0(1.5) & 4
\\
$9-$F& 19& 0.700(70) & 9.7(1.0)(0.1)(0.2) & 1.28(13)(1)(1) & 11$_{-1}^{+2}$ & 9.0(2.7) & 2
\\
${10}-$Ne& 20& 0.741(74)& 12.6(1.3)(0.2)(0.3) & 1.78(18)(2)(1) & 14.4$_{-1.4}^{+2.8}$ & 19(6) & 5
\\
& 21& 0.783(78)& 13.7(1.4)(0.2)(0.4) & 1.88(19)(2)(1) & 16$_{-2}^{+4}$ & 18(5) & 4
\\
& 22& 0.823(82)& 14.6(1.5)(0.2)(0.4) & 1.98(20)(2)(1) & 17$_{-2}^{+4}$ & 18(5) & 4
\\
${11}-$Na& 23& 0.870(87)& 18.9(1.9)(0.3)(0.6) & 2.64(26)(4)(1) & 22$_{-2}^{+4}$ & 25(8) & 2
\\
${12}-$Mg& 24& 0.915(91)& 24.6(2.5)(0.4)(0.6) & 3.46(35)(6)(2) & 28$_{-3}^{+6}$ & 38(11) & 2
\\
& 25& 0.961(96)& 25.5(2.6)(0.5)(0.8) & 3.61(36)(6)(2) & 29$_{-3}^{+6}$ & 31(9) & 3
\\
& 26& 1.01(10)& 26.2(2.6)(0.5)(0.9) & 3.75(38)(6)(2) & 30$_{-3}^{+6}$ & 33(10)& 3
\\
${13}-$Al& 27& 1.10(11)$^a$ & 34.7(3.5)(0.8)(1.2) & 4.80(48)(9)(3) & 39$_{-4}^{+8}$ & 40(12)& 2
\\
${14}-$Si& 28& 1.10(11)& 42.4(4.2)(1.1)(1.5) & 5.99(60)(12)(4) & 48$_{-5}^{+10}$ & 55(16)& 5
\\
& 29& 1.15(12)& 44.3(4.4)(1.1)(1.6) & 6.21(62)(13)(4) & 51$_{-5}^{+10}$ & 53(16) & 45
\\
& 30& 1.20(12)& 46.1(4.6)(1.2)(1.6) & 6.42(64)(13)(4) & 53$_{-5}^{+10}$ & 51(15) & 45
\\
${15}-$P& 31& 1.26(13)& 55.8(5.6)(1.6)(2.1) & 7.86(79)(18)(6) & 64$_{-6}^{+12}$ & 61(18) & 11
\\
${16}-$S& 32& 1.31(13)& 66.2(6.6)(2.0)(2.6) & 9.48(95)(24)(7) & 76$_{-7}^{+14}$ & 83(25) & 12
\\
& 34& 1.42(14)& 71.2(7.2)(2.2)(2.8) & 10.1(1.0)(0.3)(0.1) & 81$_{-7}^{+14}$ & 79(24) & 14
\\
& 36& 1.53(15)& 76.5(7.7)(2.4)(3.0) & 10.6(1.1)(0.3)(0.1) & 87$_{-8}^{+16}$ & 75(23) & 13
\\
${17}-$Cl& 35& 1.47(15)& 82.6(8.3)(2.9)(3.4) & 11.9(1.2)(0.3)(0.1) & 94$_{-9}^{+18}$ & - & -
\\
& 37& 1.58(16)& 89(9)(3)(4) & 12.6(1.3)(0.4)(0.1) & 102$_{-10}^{+20}$ & - & - 
\\
${18}-$Ar& 36& 1.53(15)& 97(10)(4)(4) & 14(1.4)(0.4)(0.1) & 111$_{-11}^{+22}$ & 118(36) & 24
\\
& 38& 1.64(16)& 104(10)(4)(5) & 15(1.5)(0.5)(0.1) & 119$_{-12}^{+24}$ & 107(32) & 24
\\
& 40& 1.75(18)& 111(11)(4)(5) & 16(1.6)(0.5)(0.1) & 127$_{-13}^{+26}$ & 126(38) & 25
\\
${19}-$K& 39& 1.70(17)& 118(12)(5)(5) & 18(1.8)(0.6)(0.2) & 136$_{-14}^{+28}$ & 119(36) & 32
\\
& 41& 1.81(18)& 126(13)(5)(6) & 18(1.8)(0.6)(0.2) & 144$_{-15}^{+30}$ & 132(40) & 28
\\
${20}-$Ca& 40& 1.75(18)& 135(14)(6)(6) & 20(2.0)(0.7)(0.2) & 155$_{-16}^{+32}$ & 142(40) & 25
\\
& 42& 1.87(19)& 143(14)(6)(7) & 21(2.1)(0.8)(0.2) & 164$_{-17}^{+34}$ & 166(50) & 29
\\
& 43& 1.93(19)& 148(15)(7)(7) & 21(2.1)(0.8)(0.2) & 169$_{-18}^{+36}$ & 145(43) & 27
\\
& 44& 2.00(20)& 152(15)(7)(7) & 22(2.2)(0.8)(0.2) & 174$_{-18}^{+36}$ & 175(52) & 26
\\
& 46& 2.12(21)& 162(16)(7)(8) & 23(2.3)(0.8)(0.2) & 185$_{-19}^{+38}$ & 156(47) & 107
\\
& 48& 2.25(22)& 173(17)(6)(7) & 24(2.4)(0.9)(0.2) & 197$_{-19}^{+38}$ & 153(46) & 26
\\
${21}-$Sc& 45& 2.06(21)& 172(17)(7)(7) & 25(2.5)(1.0)(0.2) & 197$_{-20}^{+40}$ & 203(61) & 41 
\\
${22}-$Ti& 46& 2.12(21)& 192(19)(8)(8) & 28(2.8)(1.2)(0.3) & 220$_{-22}^{+44}$ & 257(77) & 26
\\
& 47& 2.18(22)& 194(19)(8)(9) & 29(2.9)(1.2)(0.3) & 223$_{-22}^{+45}$ & 252(76) & 25
\\
& 48& 2.25(22)& 200(20)(9)(9) & 29(2.9)(1.3)(0.3) & 229$_{-24}^{+47}$ & 241(72) & 26
\\
& 49& 2.31(23)& 207(21)(9)(9) & 30(3.0)(1.3)(0.3) & 237$_{-25}^{+49}$ & 215(64) & 33
\\
& 50& 2.38(24)& 213(21)(9)(9) & 31(3.1)(1.3)(0.3) & 234$_{-25}^{+49}$ & 216(65) & 26
\\
${23}-$V& 51& 2.44(24)& 231(23)(11)(11) & 35(3.5)(1.6)(0.4) & 266$_{-28}^{+55}$ & 245(73) & 26
\\
${24}-$Cr& 50& 2.38(24)& 240(24)(12)(12) & 37(4)(2)(1) & 277$_{-29}^{+59}$ & 333(100) & 27
\\
& 52& 2.51(25)& 255(25)(13)(12) & 39(4)(2)(1) & 294$_{-31}^{+61}$ & 299(90) & 21
\\
& 53& 2.58(26)& 260(26)(13)(13) & 39(4)(2)(1) & 299$_{-32}^{+64}$ & 302(91) & 25
\\
& 54& 2.65(26)& 265(27)(13)(13) & 40(4)(2)(1) & 305$_{-33}^{+66}$ & 318(96) & 31
\\
${25}-$Mn& 55& 2.72(27)& 297(30)(16)(15) & 44(4)(2)(1) & 341$_{-37}^{+74}$ & 364(109) & 34
\\
\end{tabular}
\end{ruledtabular}
\end{table*}

\begin{table*}[ht]
\caption{\label{tab:tablexx}%
Same as in Table \ref{tab:tablex}}
\begin{ruledtabular}
\renewcommand{\arraystretch}{1.18}
\begin{tabular}{llcrrrrr}
$Z-$Element & $A$ & $\alpha_{E1}$\,(fm$^3$) & $-\Delta E_{1S}^{NP}$ &$-\Delta E_{1S}^{nP}$& Total NP & Entry in \cite{Fricke:2004} & Goal
 \\
\hline
${26}-$Fe& 54& 2.65(26)& 313(31)(17)(16) & 48(5)(3)(1) & 361$_{-39}^{+78}$ & 362(109) & 48
\\
& 56& 2.79(28)& 325(32)(18)(17) & 49(5)(3)(1) & 374$_{-41}^{+82}$ & 403(121) & 44
\\
& 57& 2.86(29)& 332(33)(19)(17) & 50(5)(3)(1) & 382$_{-42}^{+84}$ & 390(117) & 56
\\
& 58& 2.93(29)& 338(34)(19)(17) & 50(5)(3)(1) & 388$_{-43}^{+86}$ & 400(120)& 54
\\
${27}-$Co& 59& 3.00(30)& 367(37)(22)(19) & 56(6)(4)(2) & 423$_{-48}^{+95}$ & 438(131) & 50
\\
${28}-$Ni& 58& 2.93(29)& 389(39)(25)(21) & 59(6)(4)(1) & 448$_{-51}^{+103}$ & 437(131) & 46
\\
& 60& 3.07(31)& 395(40)(25)(21) & 61(6)(4)(1) & 456$_{-52}^{+104}$ & 461(138) &45
\\
& 61& 3.14(31)& 403(40)(26)(22) & 62(6)(4)(1) & 465$_{-53}^{+106}$ & 426(138) & 54
\\
& 62& 3.22(32)& 410(41)(26)(22) & 62(6)(4)(1) & 472$_{-54}^{+108}$ & 458(138) & 45
\\
& 64& 3.36(34)& 427(43)(28)(23) & 64(6)(4)(1) & 491$_{-57}^{+113}$ & 438(138) & 49
\\
${29}-$Cu& 63& 3.29(33)& 426(43)(29)(24) & 68(7)(5)(1) & 494$_{-58}^{+116}$ & 538(161) & 47
\\
& 65& 3.44(34)& 448(45)(31)(25) & 70(7)(5)(1) & 518$_{-61}^{+121}$ & 489(147) & 49
\\
${30}-$Zn& 64& 3.36(34)& 462(46)(33)(27) & 73(7)(5)(1) & 535$_{-63}^{+127}$ & 609(183) & 47
\\
& 66& 3.52(35)& 479(48)(35)(28) & 75(8)(5)(1) & 554$_{-66}^{+133}$ & 595(179) & 45
\\
& 68& 3.67(37)& 498(50)(37)(29) & 77(8)(6)(1) & 575$_{-69}^{+139}$ & 581(174) & 32
\\
& 70& 3.82(38)& 516(52)(38)(30) & 79(8)(6)(1) & 595$_{-72}^{+144}$ & 615(184) & 131
\\
${31}-$Ga& 69& 3.75(37)& 522(52)(40)(31) & 83(8)(6)(1) & 605$_{-73}^{+146}$ & 567(169) & 12
\\
& 71& 3.90(39)& 552(55)(43)(33) & 86(9)(7)(1) & 638$_{-78}^{+156}$ & 551(165) & 12
\\
${32}-$Ge& 70& 3.82(38)& 566(57)(46)(34) & 89(9)(7)(1) & 655$_{-82}^{+163}$ & 706(212) & 16
\\
& 72& 3.98(40)& 570(57)(47)(35) & 92(9)(8)(1) & 662$_{-83}^{+165}$ & 738(221) & 12
\\
& 73& 4.06(41)& 580(58)(49)(36) & 93(9)(8)(1) & 673$_{-85}^{+170}$ & 700(210) & 24
\\
& 74& 4.14(41)& 589(59)(49)(36) & 94(9)(8)(1) & 683$_{-86}^{+171}$ & 839(242) & 17
\\
& 76& 4.30(43)& 611(61)(51)(38) & 96(10)(8)(1) & 707$_{-89}^{+178}$ & 819(246) & 15
\\
${33}-$As& 75& 4.22(42)& 628(63)(54)(40) & 101(10)(9)(2) & 729$_{-93}^{+186}$ & 761(228) & 10
\\
${34}-$Se& 76& 4.30(43)& 662(66)(61)(43) & 107(11)(10)(2) & 769$_{-101}^{+202}$ & 1036(311) & 16
\\
& 77& 4.39(44)& 675(67)(62)(43) & 109(11)(10)(2) & 784$_{-102}^{+204}$ & 790(237) & 16
\\
& 78& 4.47(45)& 687(69)(63)(44) & 110(11)(10)(2) & 797$_{-104}^{+209}$ & 949(285) & 13
\\
& 80& 4.64(46)& 713(71)(65)(46) & 113(11)(10)(2) & 826$_{-108}^{+215}$ & 872(262) & 12
\\
& 82& 4.81(48)& 739(74)(68)(48) & 116(12)(11)(2) & 855$_{-113}^{+225}$ & 814(244) & 19
\\
${35}-$Br& 79& 4.55(46)& 728(73)(70)(48) & 117(12)(11)(2) & 845$_{-113}^{+226}$ & 933(280) & 17
\\
& 81& 4.72(47)& 755(76)(72)(50) & 120(12)(11)(2) & 875$_{-117}^{+234}$ & 827(248) & 20
\\
${36}-$Kr& 78& 4.47(45)& 736(73)(74)(49)&  121(12)(12)(2)&  857$_{-116}^{+232}$& 1183(355) & 40
\\
& 80& 4.64(46)& 765(77)(77)(51)&  124(12)(12)(2)&  889$_{-121}^{+243}$& 1071(321) & 40
\\
& 82& 4.81(48)& 795(80)(80)(53)&  128(13)(13)(2)&  923$_{-126}^{+253}$& 938(281) & 40
\\
& 83& 4.89(49)& 811(81)(81)(54)&  129(13)(13)(2)&  940$_{-128}^{+256}$& 936(281) & 47
\\
& 84& 4.98(50)& 825(82)(83)(55)&  131(13)(13)(2)&  956$_{-130}^{+261}$& 838(251) & 39
\\
& 86& 5.15(52)& 855(85)(85)(57)&  134(13)(13)(2)&  989$_{-134}^{+269}$& 866(260) & 34
\\
${37}-$Rb& 85& 5.06(51)& 870(87)(91)(59)&  139(14)(14)(2)&  1009$_{-140}^{+281}$& 853(256) & 10
\\
& 87& 5.24(52)& 902(90)(94)(61)&  142(14)(15)(2)&  1044$_{-145}^{+290}$& 807(242) & 14
\\
${38}-$Sr& 84& 4.98(50)& 888(89)(96)(61)&  145(14)(16)(3)&  1033$_{-146}^{+292}$& 1136(341) & 24
\\
& 86& 5.15(52)& 913(91)(99)(63)&  147(15)(16)(3)&  1060$_{-150}^{+300}$& 929(279) & 11
\\
& 87& 5.24(52)& 931(93)(101)(64)&  149(15)(16)(3)&  1080$_{-153}^{+306}$& 843(253) & 49
\\
& 88& 5.33(53)& 946(95)(103)(65)&  151(15)(16)(3)&  1097$_{-156}^{+312}$& 937(281) & 8
\\
${39}-$Y& 89& 5.42(54)& 1008(101)(114)(70)&  158(16)(18)(3)&  1166$_{-169}^{+339}$& 867(260) & 9
\\
${40}-$Zr& 90& 5.51(55)& 1059(106)(124)(74)&  166(17)(20)(3)&  1225$_{-181}^{+362}$& 975(292) & 10
\\
& 91& 5.60(56)& 1029(103)(122)(74)&  167(17)(20)(3)&  1196$_{-178}^{+356}$& 957(287) & 33
\\
& 92& 5.69(57)& 1045(104)(124)(75)&  169(17)(20)(3)&  1214$_{-180}^{+361}$& 984(295) & 13
\\
& 94& 5.87(59)& 1069(107)(128)(77)&  171(17)(20)(3)&  1240$_{-186}^{+371}$& 946(284) & 15
\\
& 96& 6.05(61)& 1095(109)(132)(78)&  174(17)(21)(3)&  1269$_{-190}^{+380}$& 966(293) & 36
\\
${41}-$Nb& 93& 5.78(58)& 1092(109)(135)(79)&  177(18)(20)(3)&  1269$_{-193}^{+385}$& 1127(338) & 16
\\

\end{tabular}
\end{ruledtabular}
\end{table*}

\end{document}